\documentclass[aip,prb,numerical,reprint]{revtex4-1}
\usepackage[normalem]{ulem}
\usepackage{gensymb,amsmath, amssymb, amsfonts, amsthm}
\usepackage{graphicx}
\usepackage{bm}
\usepackage{dcolumn}
\usepackage{lgreek}
\usepackage{enumerate}
\usepackage{placeins}

\begin{document}

\title{Current oscillations in Vanadium Dioxide:\\ evidence for electrically triggered percolation avalanches}

\author{Tom Driscoll}
\email[]{Tom.Driscoll@Duke.edu}
\affiliation{Center for Metamaterials and Integrated Plasmonics, Pratt School of Engineering, Duke University.  Durham, NC, 27708, USA}
\affiliation{Physics Department, University of California, San Diego, La Jolla, California 92093, USA}
\author{Jack Quinn}
\affiliation{Physics Department, University of California, San Diego, La Jolla, California 92093, USA}
\author{Giwan Seo}
\affiliation{School of Advanced Device Technology, University of Science and Technology (UST). Daejon, 305-350, Korea.}
\author{Yong-Wook Lee}
\affiliation{School of Electrical Engineering, Pukyong National University, Busan 608-737, South Korea.}
\author{Hyun-Tak Kim}
\affiliation{School of Advanced Device Technology, University of Science and Technology (UST). Daejon, 305-350, Korea.}
\affiliation{Creative research center of Metal-insulator transition, ETRI, Daejon 305-700, South Korea.}
\author{David R. Smith}
\affiliation{Center for Metamaterials and Integrated Plasmonics, Pratt School of Engineering, Duke University.  Durham, NC, 27708, USA}
\author{Massimiliano Di Ventra}
\author{Dimitri N. Basov}
 \affiliation{Physics Department, University of California, San Diego, La Jolla, California 92093, USA}
\date{\today}

\begin{abstract}
	In this work, we experimentally and theoretically explore voltage controlled oscillations occurring in micro-beams of vanadium dioxide.  These oscillations are a result of the reversible insulator to metal phase transition in vanadium dioxide.  Examining the structure of the observed oscillations in detail, we propose a modified percolative-avalanche model which allows for voltage-triggering.  This model captures the periodicity and waveshape of the oscillations as well as several other key features.  Importantly, our modeling shows that while temperature plays a critical role in the vanadium dioxide phase transition, electrically induced heating cannot act as the primary instigator of the oscillations in this configuration.  This realization leads us to identify electric field as the most likely candidate for driving the phase transition.

\end{abstract}

\maketitle

\section{Introduction} \label{sec:intro}
Vanadium Dioxide (VO$_2$) has been a material of prolonged scientific interest, due to the plethora of unusual properties associated with the Insulator to Metal phase Transition (IMT) occurring just above room temperature\cite{Morin1959}.  The large conductivity change ratio, combined with an accessible transition temperature and rich correlated-electron physics \cite{Qazilbash2007a,Qazilbash2008a,Liu2009a} has made this an attractive compound for many researchers.   Much attention has historically revolved around controversy over the driving physics of the phase transition; particularly whether it is a Mott transition \cite{Zylbersztejn1975,Kim2004,Rice1994b} or Pierels transition \cite{Qazilbash2007a,Wentzcovitch1994,Cavalleri2004}.  However, also of interest is the ability of the IMT to happen on ultrafast (100fs) timescales\cite{Cavalleri2001}, and the wide range of stimuli which can trigger it\cite{Rini2005,Lysenko2006}.  Along these lines, recent interest has also shifted from purely academic to industrial as well; following proposed applications ranging from optical devices \cite{Lopez2004} and hybrid-metamaterials \cite{Driscoll2008,Driscoll2009b,Dicken2009} to electronic components \cite{Kim2006,Driscoll2010a} and data storage \cite{Driscoll2009d,Pershin2011}.  With this rise of potential applications comes opportunities for new avenues of research and development, but also new challenges to satisfy the durability and flexibility that real-world devices demand \cite{Crunteanu2010}.  Understanding the role of temperature and structural transitions in various VO$_2$ phenomenon is key to pushing towards potential of applications. 

In this manuscript we take an interest in the recently reported \cite{Lee2008a,Sakai2008} phenomenon of self-sustaining oscillations in VO$_2$ bridges.  The widespread prevalence of voltage controlled oscillators in electronics makes this phenomenon an enticing candidate for devices.  It is fairly well accepted that these oscillations represent a triggering of the Insulator-to-Metal Transition (IMT), followed by a reseting Metal-to-Insulator Transition (MIT).  However, despite headway on controlling such oscillations in VO$_2$ \cite{Kim2010b,Sakai2008}, there is still debate over whether the underlying driving mechanism is thermal or electrostatic.  In literature, VO$_2$ is most often thermally triggered, and yet these oscillations appear to respond to foremost to applied voltage across the device.  The unavoidable presence of joule-heating currents through the 2-terminal device during operation, coupled with the observed sensitivity of the oscillations to device temperature\cite{Kim2010c,Crunteanu2010}, make for a contentious situation.

In our investigation we first experimentally reproduce the oscillations discovered by the authors of Ref \cite{Lee2008a}.  The use of a high-performance oscilloscope in our experiment gives us access to very fine time-resolution data which is useful in our modeling.  The details and data of our experiment are reported in Section \ref{sec:expt}. Following this, we develop a model which replicates and explains the observed waveshape in terms of electrically-triggered domains.  Our model, reported in Section \ref{sec:model}, describes a network of electrically and/or thermally triggered grains.  This model is inspired by several previously proposed models\cite{Sharoni2008,Ramirez2009}.  While these previous models also predict avalanche-like transitions under the right conditions\cite{Sharoni2008}, our our model expands on this framework to track time-dependent effects and allow possibility of a voltage-triggered phase-transition.  Alongside voltage triggering, we investigate the role of temperature in the oscillations, and importantly, we find while a voltage-driven picture replicates experimental data - thermal heating \emph{alone} is quantitatively and qualitatively unable to explain the structure of the observed oscillations.  Nevertheless, device temperature does affect oscillations, and thermal co-factors to the voltage triggering are needed to reproduce aspects of the data.  In Section \ref{sec:percolation} we discuss how the percolative transition of VO$_2$ affects the shape of the MIT, and what this means regarding effective medium within the phase-coexistence region.  In Section \ref{sec:summary} we conclude the manuscript with an overview of our results, and an outlook on possible directions for VO$_2$ research and application.

\section{Voltage controlled oscillations.} \label{sec:expt}
\begin{figure} 
\includegraphics[width=3in]{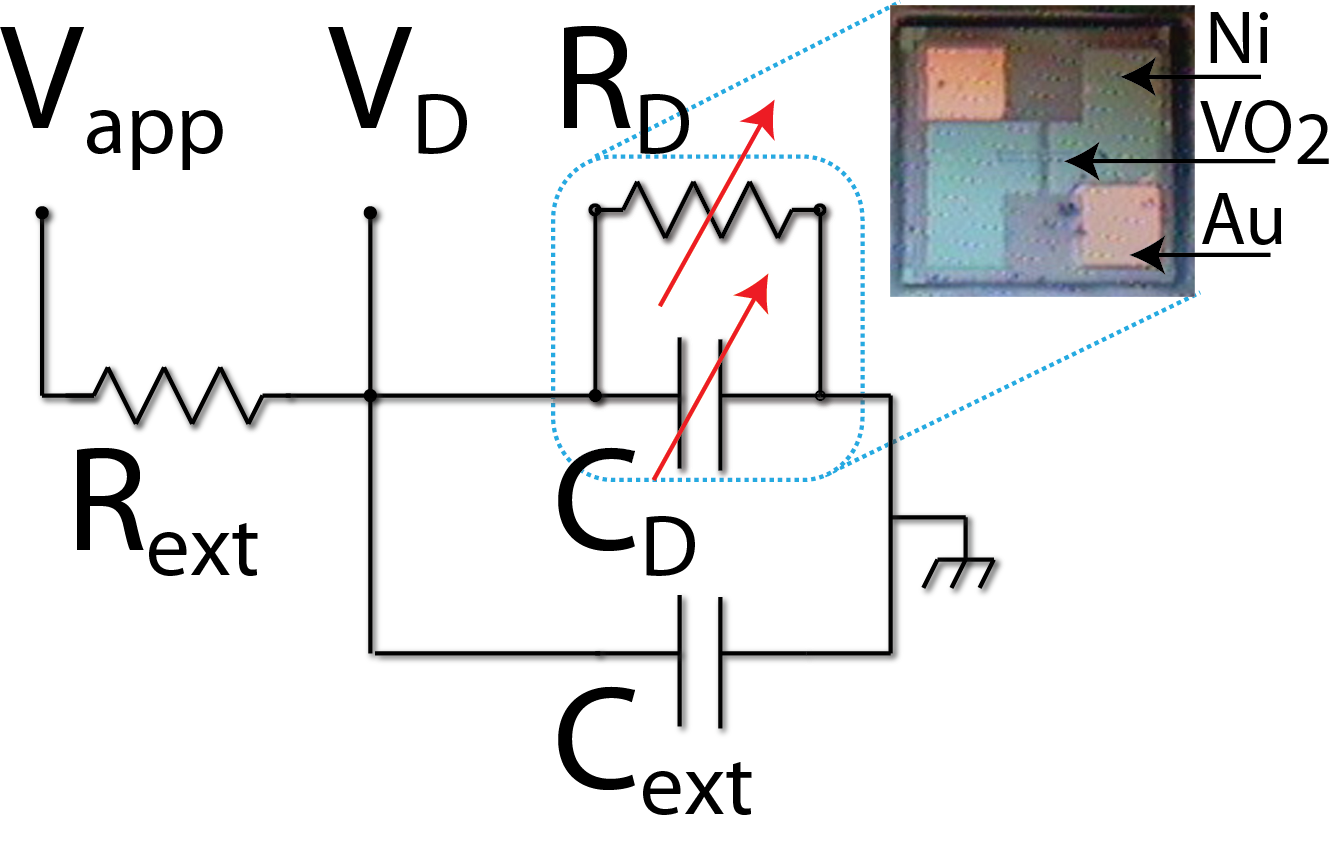}
\caption{Schematic of the circuit diagram used to reproduce oscillations, depicting the VO$_2$ device as a variable resistance and capacitance.  Inset shows optical photograph of a sample device.  The Al$_2$O$_3$ substrate is 330$\mu$m thick, and is mounted on a glass cover slip.  All experiments are performed at room temperature.}
\label{fig:RLC}
\end{figure}

\begin{figure} 
\includegraphics[width=3in]{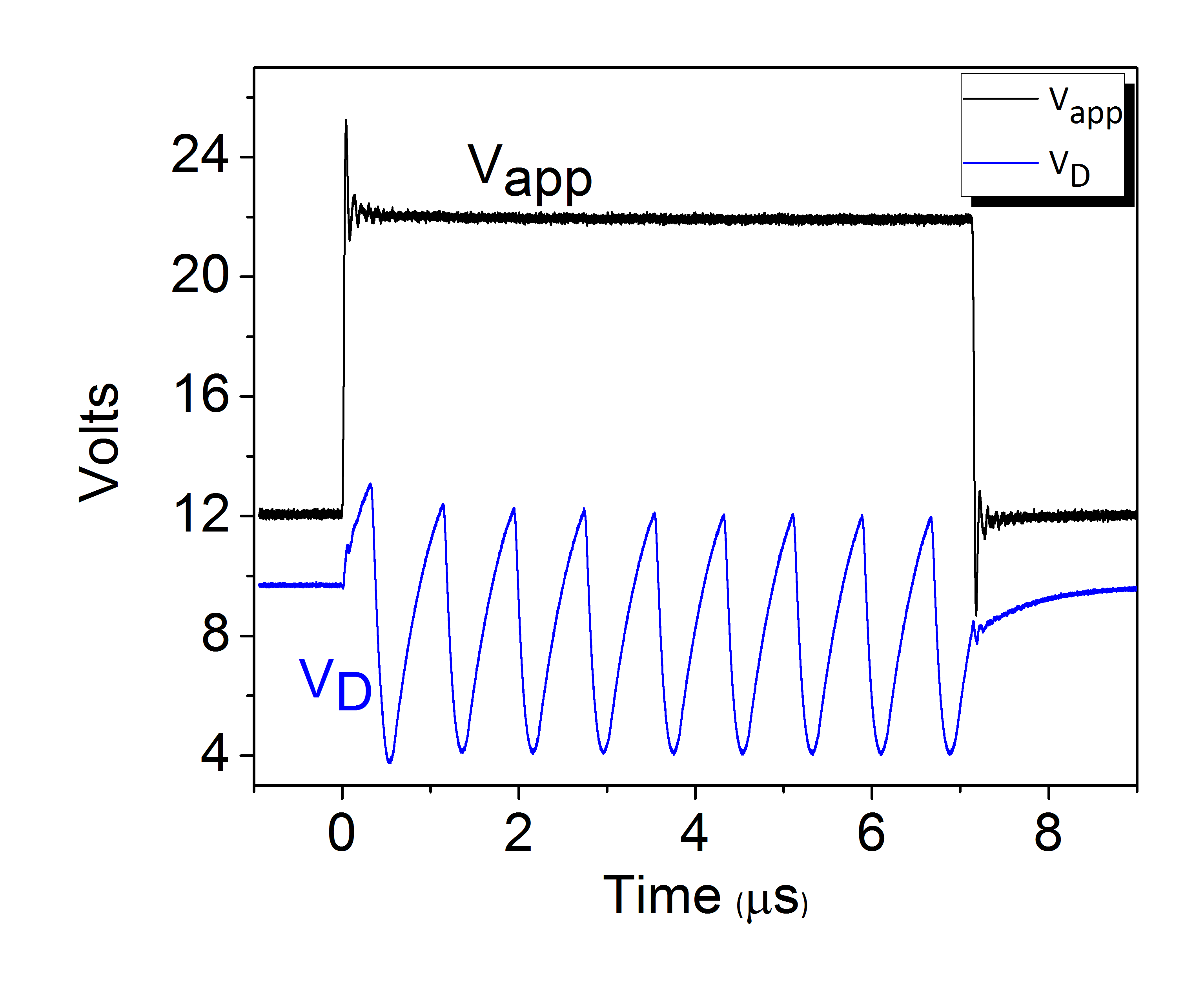}
\caption{Applied voltage and voltage across the device as a function of time.   \label{fig:osc} }
\end{figure}

Our investigation begins with replicating VO$_2$ oscillations using the procedure reported by Kim et.al \cite{Kim2010b,Lee2008a}.  A device consisting of a 10$\mu$m x 10$\mu$m VO$_2$ bridge between two large (\verb+~+400$\mu$m) metal (Ni:Au) electrodes (Figure \ref{fig:RLC}-inset ) is hooked in series with a limiting resistor (R$_{ext}$) and voltage source (V$_{app}$).  This setup is shown schematically in Figure \ref{fig:RLC}.  Although we do not intentionally add external capacitance, the presence of such C$_{ext}$ in instruments and cables is unavoidable - and should be included in the effective circuit.  The applied voltage V$_{app}$ is a transient square pulse (between 1$\mu$s to 1ms) from a Agilent function generator (model 33120A) riding on top of a constant bias voltage (V$_{bias}=12V$).  This is shown as the black trace in Figure \ref{fig:osc}, giving a peak applied voltage of 22V.  The voltage across the device (V$_D$, as shown by the blue trace in Figure \ref{fig:osc}) is monitored with a LeCroy (model wavepro 7-zi) oscilloscope, which allows for high time-resolution (40GS/s) sampling resolution even over millisecond-long pulses.

In this configuration, the VO$_2$ device functions essentially as a capacitor with a variable internal shunt resistance R$_D$.  The capacitance C$_D$ is primarily fixed by device geometry, although variations of the dielectric constant of VO$_2$ throughout the phase transition (such as have been shown in the context of memory-capacitance\cite{Driscoll2009b,DiVentra2009} and VO$_2$ hybrid-metamaterials\cite{Driscoll2008}) may have small effects - and we discuss this later in Section \ref{sec:EMA}.  The pre-pulse steady-state starting voltage is V$_D$=V$_{bias}$R$_D$/(R$_D$ + R$_{ext}$).  At the start of the pulse (t=0), V$_D$ increases, following a canonical Resistance-Capacitance (RC) charging curve.  Once V$_D$ surpasses a threshold voltage (which we will call V$_{D:IMT}$), it transitions sharply from increasing V$_D$ to decreasing.  We assign this change to an IMT event occurring in the VO$_2$, which effectively lowers the internal shunt resistance R$_D$ of the capacitor.  With lower internal resistance, the capacitor undergoes rapid discharge and V$_D$ plummets.  This discharge continues until V$_D$ reaches a lower threshold voltage (V$_{D:MIT}$ at which a second event - which we similarly assign to a a Metal-to-Insulator-Transition (MIT) - restores the high internal device resistance.  The process reverses and this sequence of events repeats; alternating charging and discharging between IMT and MIT events with a fairly stable periodicity \cite{Kim2010b}.

\section{Grain-scale model of oscillations.} \label{sec:model}
\begin{figure} 
\includegraphics[width=3in]{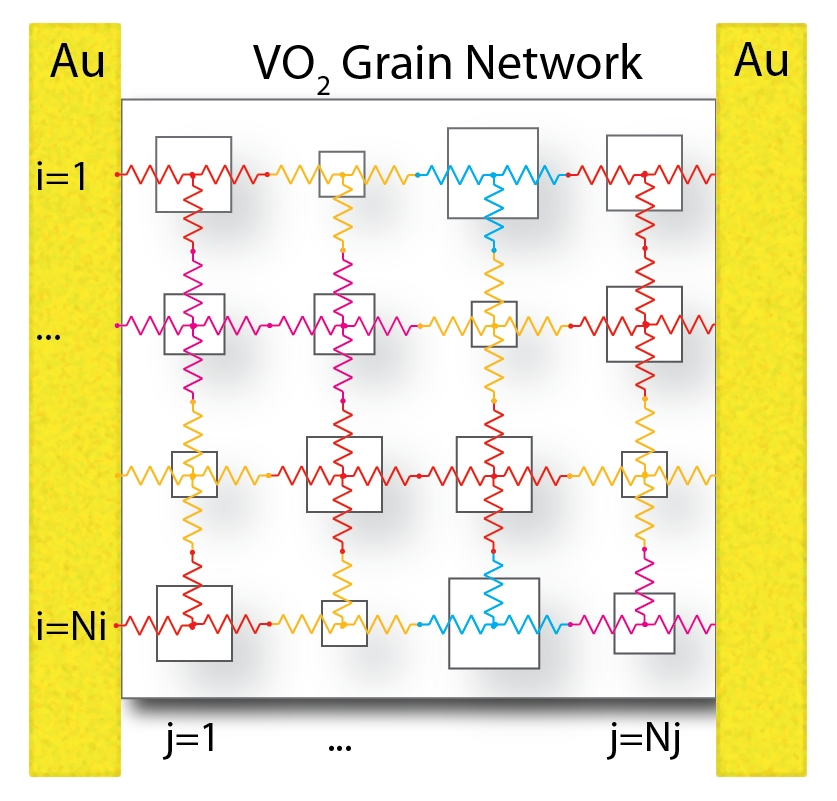}
\caption{Cartoon of the grains resistor-network.  Left and right sides are the beginning of the nickel electrodes. In the cartoon, the size of squares represent a distribution of grain sizes and resistor colors serve to highlight the corresponding different resistance.
\label{fig:network} }
\end{figure}

Our hope is that by developing a model for these observed oscillations, we may gain insight into the driving mechanism behind them.  We start with a 2D network of square VO$_2$ grains of differing size, each with total resistivity dependent on its size and its state (metal or insulator)\cite{Rozen2006,Dai2008}.  The granularity of polycrystaline VO$_2$ is well documented\cite{Pan2004,Mlyuka2006}, although the size of grains may vary considerably from one VO$_2$ preparation to another.  There is also evidence to suggest the percolation length scales for the IMT may not always coincide with the crystal granularity \cite{Kim2010d,Qazilbash2011,Frenzel2009}.  A cartoon illustrating our model arrangement is shown in Figure \ref{fig:network}.  The network consists of N$_i$xN$_j$ grains, and in our model we restrict our investigation to a 50x50 network array to keep computation time manageable. 

This grain network is placed in an external circuit containing resistance R$_{ext}$ and capacitance C$_{ext}$, and driven by V$_{app}$, as shown in Figure \ref{fig:RLC}.  The circuit differential equation (Equation \ref{eq:RLC}), is solved via Runge-Kutta time-stepping.
\begin{equation}\label{eq:RLC}
\dot{V}_D(t)=\frac{1}{(C_D+C_{ext})} \left( \frac{V_{app} - V_D}{( R_{ext})} - 		\frac{V_D}{(R_D  )}  - \dot{C_D} V_D  \right)
\end{equation}
At each time-step, we solve for the internal state of the grain network.  This includes solving a Kirchoff network problem\cite{Poklonski2006} for the voltage across each grain and the Thevenin effective circuit resistance R$_D$.  The other VO$_2$ effective circuit parameter C$_D$ is found by a differential capacitance equation (Equation \ref{eq:CD}) which can be evaluated via a self-consistent Bruggeman effective medium formulation (Equations \ref{eq:EMA1},\ref{eq:EMA2})
\begin{align} \label{eq:CD}
C_D(t) & = C_0 +  \eta \frac{\epsilon_{D} }{\epsilon_0}\\ 
0 & = f \frac{\epsilon_m - \epsilon_{D}}{\epsilon_m + 2 \epsilon_{D}} -              	(f-1) \frac{\epsilon_i - \epsilon_{D}}{\epsilon_i + 2\epsilon_{D}} 		 	\label{eq:EMA1} \\
f &=\frac{\sum^{ij} X^{ij}}{N_i N_j} \label{eq:EMA2}.
\end{align}
Superscript \emph{i} and \emph{j} are row and column indices for the grains (running to N$_i$ and N$_j$ total). The binary matrix X$^{ij}$ = 1 if the grain \emph{i,j} is metal and 0 if insulator.    $\eta$ is a capacitive fractional-fields factor (the proportion of the device capacitance which involves the VO$_2$ dielectric), which is found via finite element simulation using the COMSOL commercial code package.  C$_0$ is a geometrical capacitance which is determined empirically, fitting 1/RC to the capacitive charging curve.  External circuit parameters such as R$_{ext}$ and V$_{app}$ are taken directly from experimental values.   The extrema values for R$_D$(metal) and R$_D$(insulator) are taken from temperature data.

\subsection{Thermal triggering} \label{sec:thermal}
\begin{figure}  
\includegraphics[width=3in]{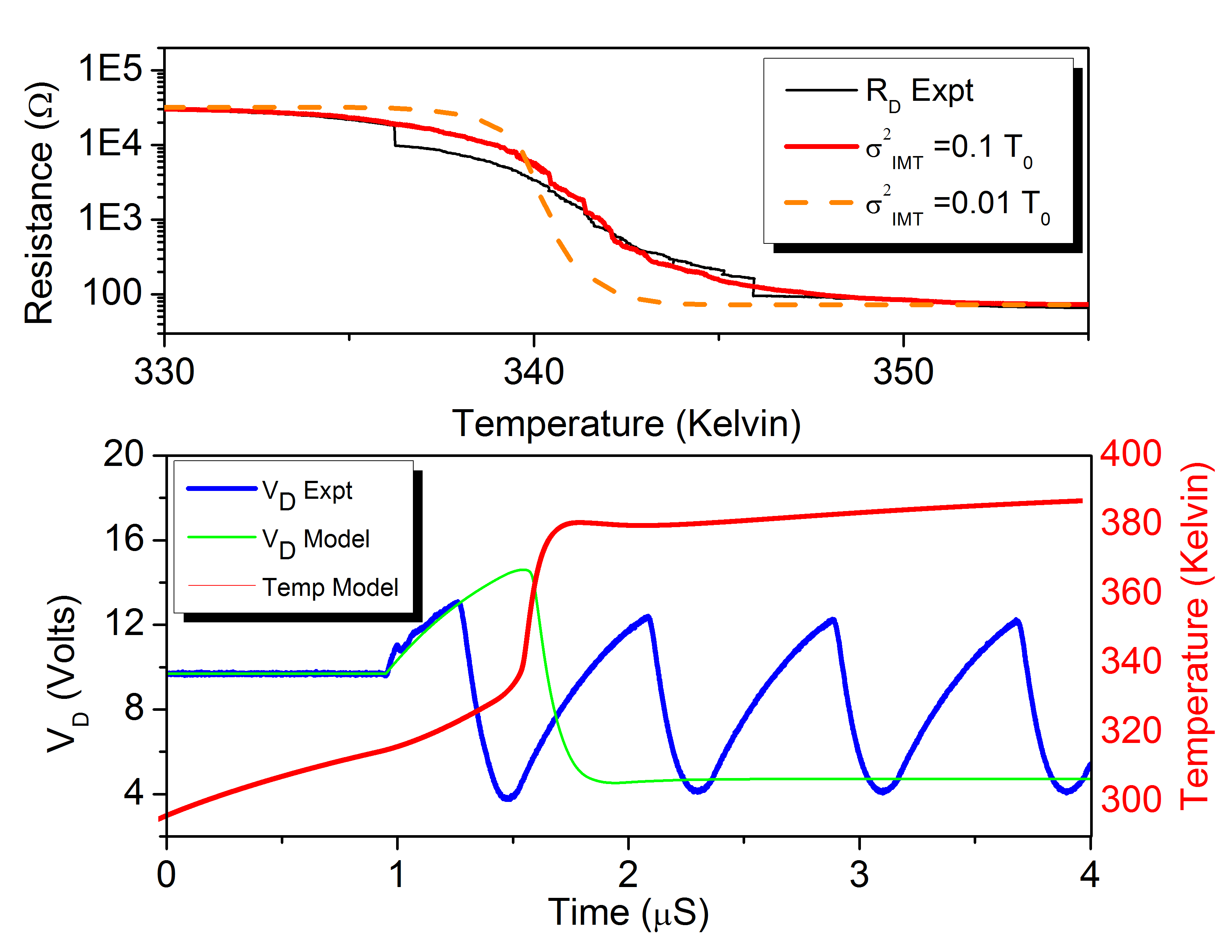}
\caption{a) Experimental resistance of our device as a function of temperature R$_D$(T) (black).  R$_D$(T) results from our model are overlaid for a best-fit value of variance: $\sigma_{IMT}^2=0.1*T_0$ (see Equation \ref{eq:Tdist}), which replicates the observed temperature-driven IMT fairly well.  Also shown is a poor-fit result for $\sigma_{IMT}^2=0.01*T_0$, given to show the effect of this parameter.   b) Attempt to replicate oscillations using thermal-only triggering, with values of $P(T_{IMT}^{ij})$ from above.  We plot V$_D$ from our model (green) overlaid on experimental V$_D$ (blue).  We observe no oscillations, only a single IMT event.  We also plot the average VO$_2$ temperature T$_D$ from our model, illustrating a runaway heating behavior that precludes oscillations.
\label{fig:RT} }
\end{figure}

The temperature driven IMT-MIT has been investigated in great detail \cite, and we ground our model using experimental data giving resistance as a function of temperature R(T) through the phase transition.  This data is shown in Figure \ref{fig:RT}a, and displays the characteristic sharp change in resistivity around 345K.  

Then, in a procedure similar to previous works\cite{Sharoni2008,Kim2004}, we assume each grain will undergo an IMT in response a "high-threshold" temperature T$_{IMT}^{ij}$, and a MIT at a low-threshold T$_{MIT}^{ij}$.   
\begin{align}
R^{ij} &= R_{met} \quad if (T^{ij} > T_{IMT}^{ij}) \\
           &= R_{ins}   \quad if (T^{ij} < T_{MIT}^{ij})
\end{align}
We assign a stochastic distribution to the values of $T_{IMT}^{ij}$ and $T_{MIT}^{ij}$ throughout the network, following the Gaussian form
\begin{align}
\label{eq:Tdist}
P(T_{IMT}^{ij})=e^{-\tfrac{(T_{IMT}^{ij}-T_{0_{IMT}})}{2\sigma_{IMT}^2}} \\
P(T_{MIT}^{ij})=e^{-\tfrac{(T_{MIT}^{ij}-T_{0_{MIT}})}{2\sigma_{MIT}^2}}.
\end{align}
The values of T$_0$ and $\sigma^2$ are fit to the experimental R(T) data shown in Figure \ref{fig:RT}a.  From this, we find a variance of $T_0=340^o$K and $\sigma_T^2$=.1*T$_0$ reproduces the shape of the IMT fairly well.  This fit gives us confidence that we understand the thermal response of VO$_2$, even though our thermal model is simple compared to some previous treatments. \cite{}  

Using this temperature-only-triggered model, we attempt to reproduce the oscillations shown in Figure \ref{fig:osc}.  Our model tracks the power dissipated in each grain and employs a finite-element method to solve for the grain and substrate temperatures as a function of time.  Material thermal parameters are taken from literature, and the enthalpy of phase-transition for VO$_2$ is included.  Figure \ref{fig:RT}b re-plots our experimental $V_D$ oscillations in blue, and the model results in green.  The result is striking: although we can track $V_D$ for a short while, we are unable to produce any oscillatory phenomena.  

Generalizing the behavior that prohibits oscillations, thermal-initiated IMTs exhibit a \emph{runaway} behavior rather than the self-stabilizing oscillatory nature seen in Figure \ref{fig:osc}.  This is apparent in the average VO$_2$ grain temperature plot (red) of \ref{fig:RT}b.  Note that although individual different grains may attain different temperatures over the course of oscillations, such gradients equalize quickly within the network.  Average temperature remains a fairly accurate and easily visualized metric of the VO$_2$ oscillation thermodynamics.  As the thermally-triggered IMT occurs, VO$_2$ temperature skyrockets even while V$_D$ discharges.  Once the temperature of the discharged device has surpassed $T_{0_{IMT}}$, there is no way for it to cool thereafter.  Power dissipation and temperature both settle towards steady-state with the device firmly in the metallic state.  The causes of this process will become more apparent as we discuss thermal dynamics in Section \ref{sec:thermal2}.

\subsection{Electrical triggering} \label{sec:electrical}

With the failure of a temperature-\emph{only}-triggering model to produce oscillations, and following insights from previous work on voltage-induced effects in VO$_2$,\cite{Sharoni2008,Ruzmetov2009,Stefanovich2000a,Lee2008a} we now introduce an electric-field driven transition.  To do this, we also assign a \emph{voltage drop} at which each grain undergoes phase transition V$_{IMT}^{ij}$ and V$_{MIT}^{ij}$.  We again use random values from a normal distribution as we did for T$_{IMT}^{ij}$ and T$_{MIT}^{ij}$.  
\begin{align}
\label{eq:Vdist}
P(V_{IMT}^{ij})=e^{-\tfrac{(V_{IMT}^{ij}-V_{0_{IMT}})}{2\sigma_{IMT}^2}} \\
P(V_{MIT}^{ij})=e^{-\tfrac{(V_{MIT}^{ij}-V_{0_{MIT}})}{2\sigma_{MIT}^2}}.
\end{align}
We lack direct data to which to fit these distributions (as we did in Figure \ref{fig:RT}a).  Thus, we retain the same value for the variance found in above, $\sigma_V^2$=0.1*V$_0$.  From an energetics perspective, this makes a great deal of sense - both distributions are surely tied to the same underlying Mott-physics.  V$_0$ remains a fitting parameter in our model.  

\begin{figure}
\includegraphics[width=3in]{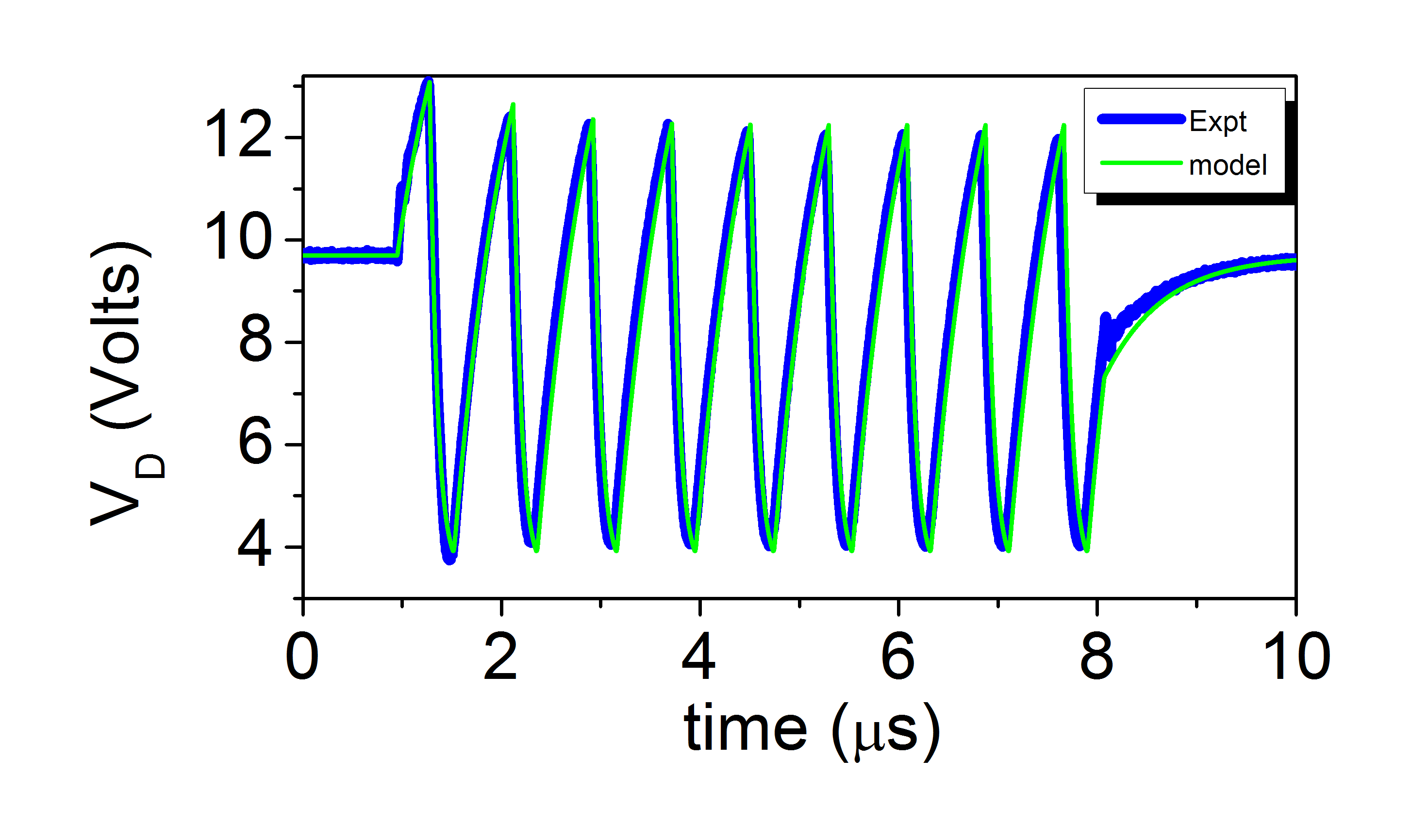}
\caption{The experimental data from Figure \ref{fig:osc} replotted (blue) with numerical results from our model overlaid (green).  This electrical-and-thermal triggering model replicates the experimental data quite well, tracking the oscillation periodicity and producing similar transitions at both V$_{D:IMT}$ and V$_{D:MIT}$.  )
\label{fig:oscmodel} }
\end{figure}

As we have not removed possible thermal triggering, the conditions for grain transition can now be stated as:
\begin{align}
R^{ij} &= R_{met} \quad if (( V^{ij}>V^{ij}_{IMT})\mbox{ OR }(T^{ij} > T_{IMT})) \\
           &= R_{ins}   \quad if (( V^{ij}<V^{ij}_{MIT})\mbox{ AND }(T^{ij} < T_{MIT}))
\end{align}
This combined triggering criterion reproduces both the waveshape and periodicity of the experimentally observed oscillations quite well.  In Figure \ref{fig:oscmodel}, we re-plot the experimental data from Figure \ref{fig:osc} in blue, and numerical results from our model are shown overlaid in green.  

\subsection{Voltage-temperature dependence} \label{sec:VT}

\begin{figure}
\includegraphics[width=3in]{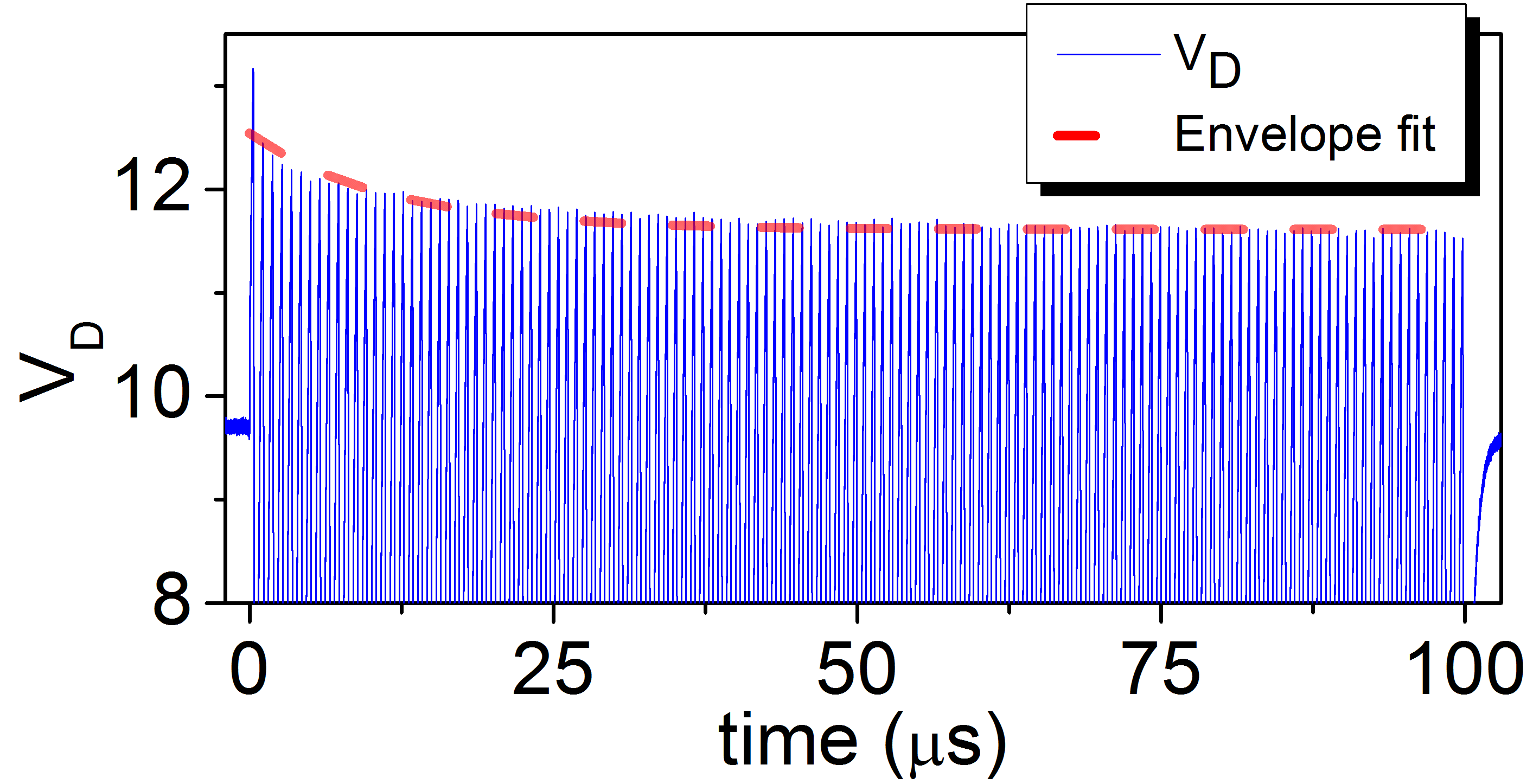}
\caption{Experimental data (blue) for oscillations over a long 100$\mu$s V$_{app}$ pulse.  We observe a clear decay envelope to the amplitude of V$_{IMT}$ in the oscillations.  For clarity, the bottom half of the oscillations are omitted as we observe no enveloping here.  To show this decay is a thermalization envelope caused by device heating, we also plot the temperature dependent V$_{0_{IMT}}(T)$ from equation \ref{eq:RofT} (red).  
\label{fig:oscenv} }
\end{figure}

Although section \ref{sec:electrical} demonstrated voltage is the primary trigger, device temperature still plays a role in oscillations.  There is a known dependence of oscillation amplitude on device temperature\cite{Kim2010c}.  If we look carefully at the data in Figure \ref{fig:oscmodel}, we notice a subtle decay envelope to the amplitude of the V$_{IMT}$ oscillation peaks.  We believe this envelope is caused by a thermalization of the device on a multi-oscillation timescale.  To accommodate this, we include a temperature-dependence to the IMT transition voltage as
\begin{equation}\label{eq:RofT}
V_{IMT}^{ij}(T)\rightarrow(\kappa(T-T_0)+1)V_{IMT}^{ij} 
\end{equation}
where T$_0$=295$^o$K. $\kappa$ is a liner temperature coefficient, the fitting of which we discuss below.  Without including this voltage-temperature interplay, our model quickly loses sync with the experimental data over the course of several oscillation periods.  This thermalization envelope is most clearly observed over a long pluse, and experimental data (solid blue) for a 100$\mu$s pulse is shown in Figure \ref{fig:oscenv}.  For clarity in this figure, the bottom half of the oscillations is omitted from view (interestingly, we observe no envelope of V$_{MIT}$).  

Using the data from Figure \ref{fig:oscenv} combined with our thermal-finite element model, we can fit a value for $\kappa$.  Overlaid on the experimental data (dashed line) is this fit, which gives $\kappa=-8.3$x$10^{-3}$.   This value of $\kappa$ is used in our oscillation model.

Explanation of this a long time-scale thermalization is straightforward.  Although locally the VO$_2$ film may heat or cool quite quickly in response to current through its volume, the 500$\mu$m thick sapphire substrate is comparatively massive.  The large thermal inertia of the substrate smooths out oscillatory heating away from the film; and when heated only from the top the substrate can require tens of micro-seconds to reach a steady-state temperature gradient.  

\subsection{Temperature dynamics} \label{sec:thermal2}

In the above Section \ref{sec:VT} we have discussed the critical interplay between temperature and voltage-triggering, showing how a long multi-oscillation timescale thermalization envelopes the oscillation amplitude. In this section, we look closer at the temperature evolution on the timescale of the oscillation period.  

\begin{figure}
\includegraphics[width=3in]{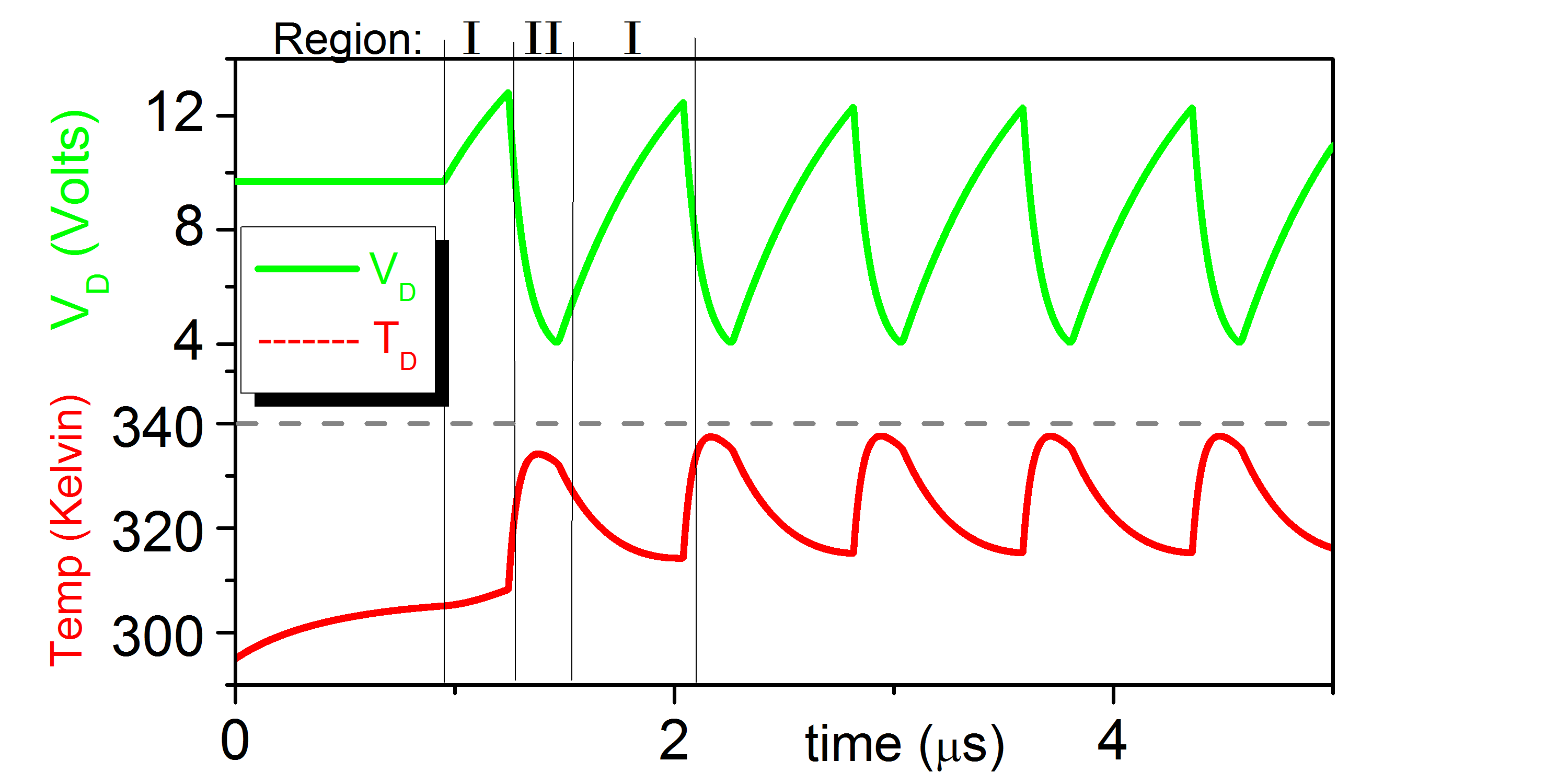}
\caption{Model results of temperature dynamics during V$_D$ oscillations.  The horizontal dashed line at T$_{IMT}$=340$^o$K shows mean transition temperature, and demonstrates that the average temperature reached during oscillations falls short of what's needed to drive the IMT transition.  We also divide the oscillation into insulating/charging regions({\bf I}) and metallic/discharge regions({\bf II}).  This draws attention to the sharp \emph{increase} in temperature which occurs only \emph{after} the V$_{IMT}$ trigger, qualitatively discrediting a thermal-driving picture for oscillations.
\label{fig:osctemp} }
\end{figure}

To begin, in Figure \ref{fig:osctemp} we plot the average grain temperature (red) during oscillations along with $V_D$ (green) - both from our model.  Looking at Figure \ref{fig:osctemp} quantitatively, we notice that the average temperature reached is not sufficient to \emph{trigger} oscillations.  Although the peak average temperature during the discharge cycle of the oscillation comes close to reaching T$_{IMT}$, a thermal-driving event would have to occur at-or-before the peak in V$_D$.  The model results indicate that the device is well below T$_{IMT}$ when V$_{D:IMT}$ is reached.  The temperature range in Figure \ref{fig:osctemp} agrees fairly well with previous numerical work on the subject \cite{Gopalakrishnan2009a}, and our own investigations using commercial finite element package COMSOL.  Compared to COMSOL, our home-grown finite-element code over-estimates temperatures reached - perhaps due to difficulty in modeling all of the spatially massive substrate (this is not an issue for commercial packages such as COMSOL).  However, commercial codes cannot be run from within Runge-Kutta time-stepping of equation \ref{eq:RLC}, and our code thus critically allows us to calculate temperature dynamics \emph{during} oscillations.

Although a quantitative argument for non-thermal triggering seems compelling, we are acutely aware that precisely solving for temperature can be difficult in such nanoscale systems. Material properties can differ from published bulk values, and interface effects can dominate transport and heating\cite{DiVentra2008}.  As we look closely though, the power dissipation in the device also appears \emph{qualitatively} unfit to explain the oscillations.  A thermally-driven transition would have to follow the logical sequence: 

\begin{enumerate}[I]
\item The insulating device heats with applied V$_D$ until it reaches T$_{IMT}$, where it undergoes an IMT (becoming metallic).  
\item The metallic device discharges its stored capacitive energy through the its own volume - cooling as it discharges - until it reaches T$_{MIT}$ where it undergoes MIT (becoming insulating).  
\item The process repeats.  
\end{enumerate}

However, Figure \ref{fig:osctemp} illustrates that region {\bf II} which occurs after V$_{D:IMT}$ is a region of \emph{maximum} power dissipation; a region of heating not cooling.  A simple ohms-law argument explains:  Just before and just after the IMT, V$_D$ is approximately V$_{D:IMT}$.  However, the resistance R$_D$ has changed by a factor of 10, and thus the power dissipated (P=V$^2_{D:IMT}$/R$_D$) is substantially greater during region {\bf II} than during region {\bf I}.  Thus, we come to the conclusion that a purely-thermal explanation for the oscillations is \emph{qualitatively} as well as quantitatively mismatched to experimental data.

Summarizing Section \ref{sec:model}, we have identified voltage as a key player in triggering observed oscillations on the grounds of several thermal arguments.  One interesting question then is whether electrostatic voltage may also trigger the also IMT in a current-free (ie. FET) configuration.  The joule-heating present in our two-terminal device complicates matters, in light of the voltage/temperature interplay identified in Section \ref{sec:VT}.  Previous work has suggested such electrostatic switching can exist,\cite{Chudnovskiy2002a,Qazilbash2008,Kim2004} although these early results await further confirmation.  Exploring the phase-space defined by the interplay of temperature, electrostatic field, and current in these VO$_2$ oscillations may reveal information about the correlated electron dynamics and energy scales associated with the Mott-transition.

\section{Percolation }\label{sec:percolation}

In this section, we go into further detail on the mechanisms of the IMT and MIT transitions.  Polycrystalline VO$_2$ is known to exhibit percolative behavior during phase transition\cite{Qazilbash2007a,Qazilbash2011,Sharoni2008,Wu2011}, and this has interesting effects on a voltage-triggered transition.  Using the model from Section \ref{sec:model} which accurately predicts the observed electrical oscillations, we attempt to gain insight on several of the internal processes during oscillatory events.

\subsection{Percolative avalanche driven oscillations} \label{sec:avalanche}

In several previous works\cite{Sharoni2008,Ramirez2009}, avalanche-like MIT and IMT transitions have been observed under the right conditions.  The immediacy of the observed change from charging to discharging in our oscillations leads us to suspect similar avalanche behavior in our electrically-driven device.  Examining the details of our model, we see that the voltage drop across any grain in the network (see Figure \ref{fig:network}) is proportional to the resistance of the grain.  During the charging cycle of the waveform voltage across the entire device (V$_D$) increases, and V$^{ij}$ across each grain does as well.  This charging continues until one "unlucky" grain hits its V$_{IMT}^{ij}$ first.  Because the grains receive a stochastic distribution for V$^{ij}_{IMT}$ (see Equation \ref{eq:Vdist}), this can be a random grain anywhere in the network.   In experiments, it is seen that the phase-transition is often seeded at particular places such as defects or boundaries.  

\begin{figure*}
\includegraphics[width=7in]{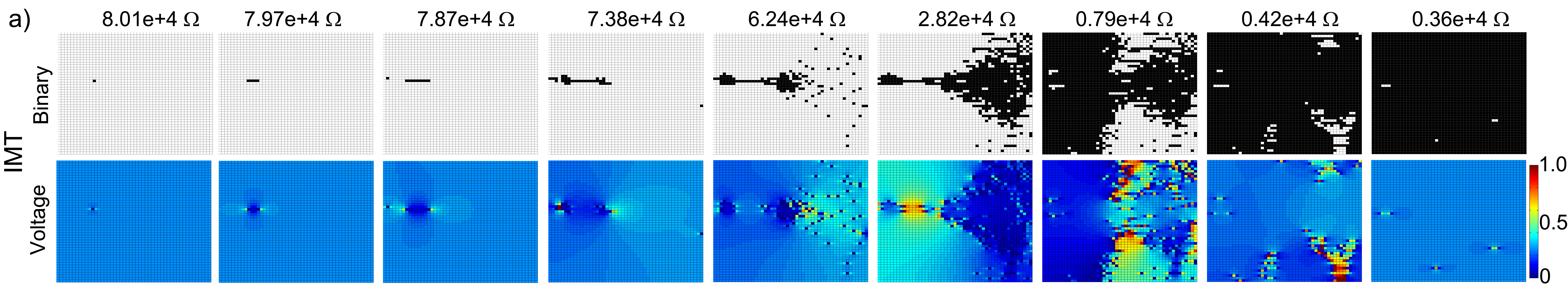}
\includegraphics[width=7in]{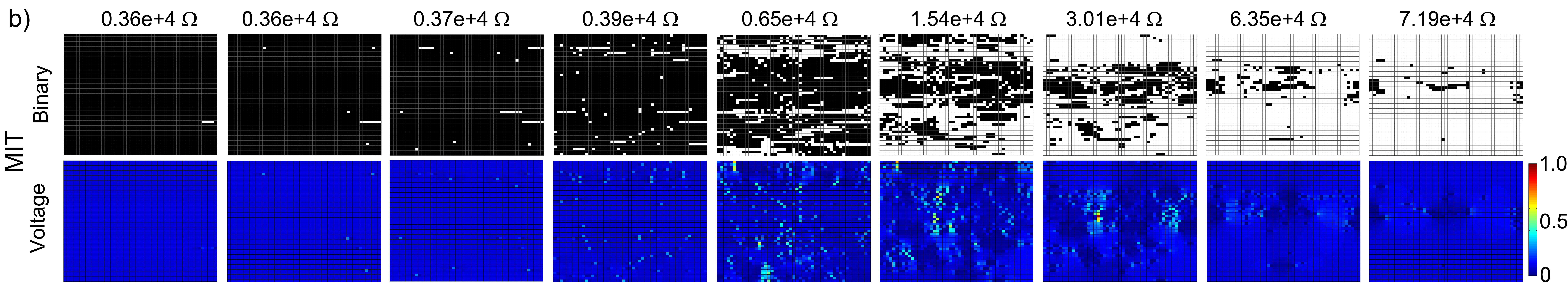}
\caption{Step-by-step depiction of the avalanche-like transition for a 50x50 grain network.  The time of each frame increases from left to right.  a) shows the IMT transition occurring at $V_{D:IMT}$, giving a bi-color plot (top) indicating whether each grain is metallic or insulating and the the voltage (bottom) across each grain in the network.   b) shows the same plots for the MIT transition occurring at $V_{D:MIT}$.  
\label{fig:PercOsc} }
\end{figure*}

The unlucky grain that first hits its IMT trigger condition undergoes an IMT.  Once this grain becomes metallic, it supports a lower voltage drop (R$_{ins}$/R$_{met} \approx$ 20) - which shifts much of its voltage burden to neighboring grains.  The neighboring grains in turn become increasingly likely to undergo their own IMT events.  The IMT spreads across the entire sample in an avalanche-like manner.  This process is depicted in Figure \ref{fig:PercOsc}a for a network of 50x50 grains.  The upper sequence of black and white frames shows whether each grain is insulating (white) or metallic (black).  The lower color frames depict the voltage drop across each grain.  The neighbor-neighbor grain interaction, mediated by voltage drop is quite evident.  

As is common in percolative systems, the Thevenin resistance R$_D$ of the network is quite sensitive to the spatial distribution of triggered grains.  The device resistivity R$_D$ is shown above each frame in the sequence of Figure \ref{fig:PercOsc}, and we see the largest drop occurs in frames \verb+#+6 and 7, where the percolation path is completed from left to right.  Once a conducting path forms $R_D$ plummets, and V$_D$ begins to drop as the device discharges.  

During discharge, there comes a point where V$_D$ drops far enough that a similar process happens in reverse.  This MIT is depicted in Figure \ref{fig:PercOsc}b.  However, the MIT process is \emph{not} exactly the reverse of the IMT.  The equations governing resistors in parallel tend to "favor" low-resistances in the following manner:  Decreasing the value of one resistor (in a parallel network) lowers the Thevenin resistance significantly, but raising the value of a single resistor has only a little effect on the Thevenin resistance.  For this reason, the avalanche-like behavior observed in Figure \ref{fig:PercOsc}a is not seen in Figure \ref{fig:PercOsc}b.  Instead, the process much more closely resembles random percolation, with only moderate neighbor-neighbor interaction.  This difference in the mechanisms between IMT and MIT may explain the difference in sharpness of the transitions at V$_{D_{IMT}}$ and V$_{D_{IMT}}$ observed in experiments.\cite{Kim2010b}

\subsection{Effective medium effects} \label{sec:EMA}

The percolative nature of the VO$_2$ phase transition allows for an inhomogeneous intermediate state where both metallic and insulating VO$_2$ coexist.  The dielectric constants of metal and insulating phase VO$_2$ are distinct; and when both phases can be present in a composite it leads to interesting properties.  The average response of the inhomogeneous sample is described by an effective medium, and can have radically different values than either.  This leads to quite interesting and novel effects.  For example, the inhomogeneity \cite{Qazilbash2007a,Frenzel2009} of polycrystalline VO$_2$ mid-transition is responsible for observed memristance \cite{Driscoll2009d} and memory-capacitance \cite{Driscoll2009b,DiVentra2009}.  

\begin{figure}
\includegraphics[width=3in]{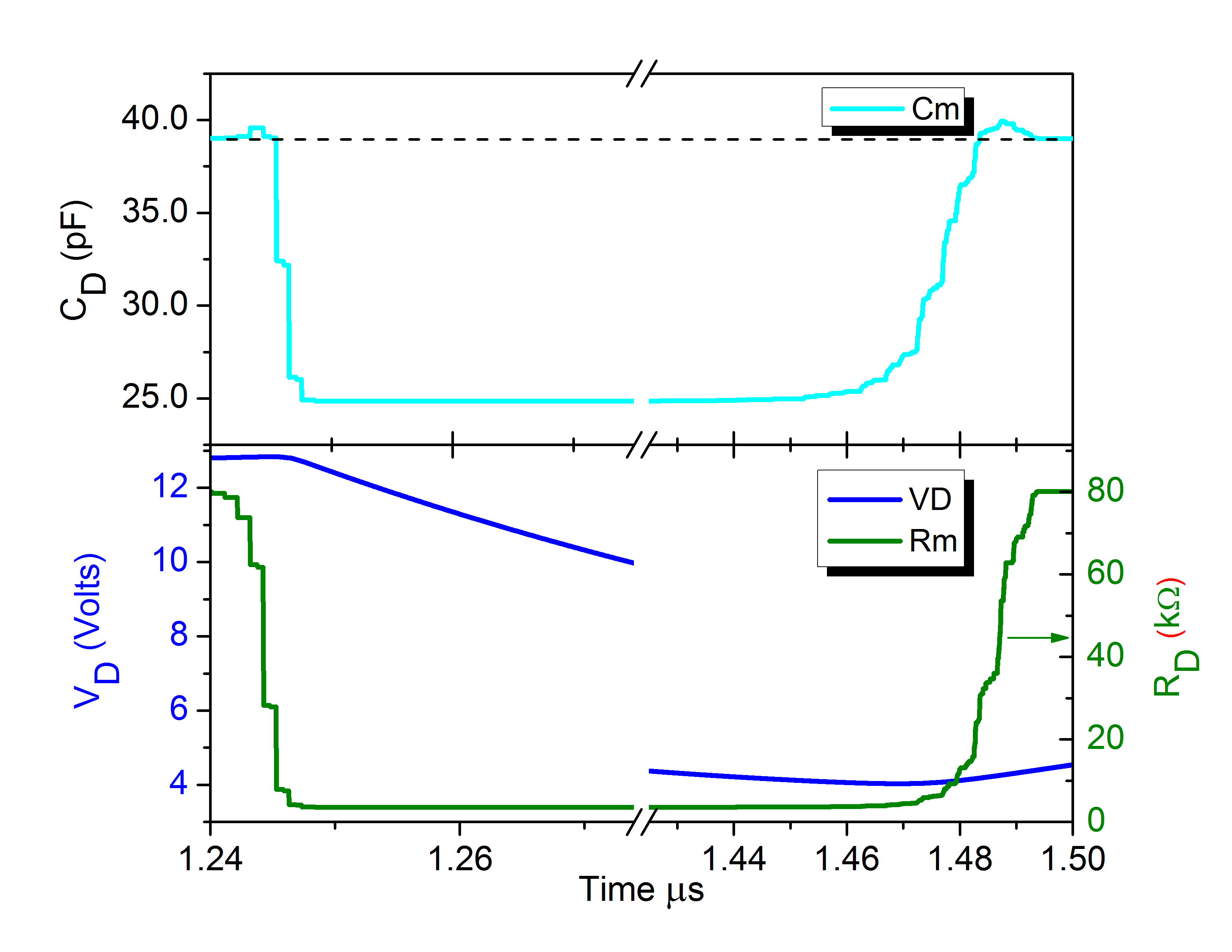}
\caption{Model values for C$_D$ plotted with V$_D$ and R$_D$ over one oscillation period.  Effective medium within the VO$_2$ (see Equation \ref{eq:CD}) causes a brief spike in C$_D$ at immediately near the IMT and MIT transitions. As seen, this spike effect is quite small compared to the overall change of both C$_D$ and R$_D$, and exists only for a very brief fraction of the oscillation period.  The jagged shape of the model curves for R$_D$ and C$_D$ reveal the small jumps typical of percolation discussed in Section \ref{sec:avalanche}
\label{fig:OscCD} }
\end{figure}

The same memory-capacitance as reported in \cite{Driscoll2009b} has previously been attributed as playing a key role in voltage-controlled oscillations in VO$_2$\cite{Kim2010b}.  This is a question we are situated to investigate in more depth using our model.  To look closely at the effects of capacitance on the oscillations, we focus attention on a single IMT transition event.  In Figure \ref{fig:OscCD}, we plot C$_D$(t) (as calculated from equations \ref{eq:CD}-\ref{eq:EMA2}), along with the familiar V$_D$ and R$_D$.   

Looking at Figure \ref{fig:OscCD}, as VO$_2$ transitions from insulating to metallic at the IMT, R$_D$ drops monotonically to its metallic-state value.  The capacitance C$_D$, however, briefly increases before also decreasing to its metallic-state value.  This increase is due to the coexistence of metallic and insulating grains, and is predicted by effective medium (Equation \ref{eq:EMA2}).  However, as Figure \ref{fig:OscCD} shows, the increase in C$_D$ is a small effect, and is contained to a short timespan near the start of the IMT.  This leads us to believe the effective medium behavior of C$_D$ has only a minor influence on the shape of the oscillations, and is not a primary driver.  We observe the same increase in C$_D$ at the MIT transition edge, but it also is too small and short-lived an effect to bear responsibility for the oscillations.  

\FloatBarrier

\section{Summary} \label{sec:summary}

In this work, we have discussed the thermal and electrical driving mechanisms behind observed oscillations occurring in VO$_2$ films.  In addition to experimentally confirming the oscillations reported by Y.W. Lee et.al.\cite{Lee2008a}, we have compiled a numerical model which is able to replicate and explain these oscillations in terms of an voltage-triggered Insulator-to-Metal phase-transition.  Temperature is known to trigger the IMT in VO$_2$, and temperature plays some role in the shape of the oscillations.  However, we find that temperature-only triggering cannot explain oscillations.  This result is likely of great importance for applications of these oscillations, as repeated thermal-cycling typically appreciably shortens device lifetime.

One question that remains unaddressed is the role, if any, of the structural phase transition in an electric-field triggered transition.  The temperature driven IMT in VO$_2$ exhibits a structural transition that happens concurrent with the electronic reconfiguration.  However, there is evidence \cite{Qazilbash2011,Qazilbash2007a,Kim2004,Arcangeletti2007,Wei2009,Yao2010a,Kim2007,Kim2008b,Kim2005} to suggest the electronic and structural transitions are not necessarily linked, but merely overlaid.  This structural electronic decoupling suggests the electronic correlations in VO$_2$ play an important role in the IMT phase transition.

Studies which probe the crystal structure simultaneous with these oscillations have not yet been reported - likely because the time and length scales associated with the VO$_2$ oscillator devices greatly complicate experiments such as x-ray diffraction.  However, as mentioned in the introduction (Section \ref{sec:intro}) the question of whether or not the structural transition occurs has great implications about the longevity of these devices.  In many applications, devices could easily be expected to perform 10$^{12}$ to 10$^{14}$ oscillation events over their lifetime, a likely impossibility if crystallographic changes are occurring.  An evident goal for the new future is to experimentally investigate the existence of structural transition in oscillations in the near future.

\section{Acknowledgments}
T.D. acknowledges support from an IC postdoctoral fellowship.  M.D. acknowledges partial support from NSF.  This research is supported by AFOSR and ETRI

\section{References}

\bibliographystyle{unsrtnat}
\bibliography{vo2osc}

\begin{thebibliography}{49}
\providecommand{\natexlab}[1]{#1}
\providecommand{\url}[1]{\texttt{#1}}
\expandafter\ifx\csname urlstyle\endcsname\relax
  \providecommand{\doi}[1]{doi: #1}\else
  \providecommand{\doi}{doi: \begingroup \urlstyle{rm}\Url}\fi

\bibitem[Morin(1959)]{Morin1959}
F~J Morin.
\newblock {Oxides which show a metal-to-insulator transition at the neel
  temperature}.
\newblock \emph{Physical Review Letters}, 3\penalty0 (1):\penalty0 34--36,
  1959.

\bibitem[Qazilbash et~al.(2007)Qazilbash, Brehm, Chae, Ho, Andreev, Kim, Yun,
  Balatsky, Maple, Keilmann, Kim, and Basov]{Qazilbash2007a}
Mumtaz~M. Qazilbash, Markus Brehm, Byung-Gyu Chae, P.-C. Ho, Gregory~O.
  Andreev, Bong-Jun Kim, Sun~Jin Yun, A.V. Balatsky, M.B. Maple, Fritz
  Keilmann, Hyun-Tak Kim, and Dimitri~N. Basov.
\newblock {Mott Transition in VO2 revealed by Infrared Spectroscopy and
  Nano-Imaging}.
\newblock \emph{Science}, 318\penalty0 (December):\penalty0 1750--1753, 2007.

\bibitem[Qazilbash et~al.(2008{\natexlab{a}})Qazilbash, Schafgans, Burch, Yun,
  Chae, Kim, Kim, and Basov]{Qazilbash2008a}
Mumtaz~M. Qazilbash, AA~Schafgans, KS~Burch, SJ~Yun, BG~Chae, BJ~Kim, Hyun-Tak
  Kim, and DN~Basov.
\newblock {Electrodynamics of the vanadium oxides VO2 and V2O3}.
\newblock \emph{Physical Review B}, 77\penalty0 (115121), 2008{\natexlab{a}}.
\newblock \doi{10.1103/PhysRevB.77.115121}.
\newblock URL \url{http://arxiv.org/abs/0803.2739}.

\bibitem[Liu et~al.(2009)Liu, Pardo, Qazilbash, Yun, Chae, Kim, Basov, and
  Averitt]{Liu2009a}
M.~Liu, B.~Pardo, MM~Qazilbash, S.J. Yun, BG~Chae, BJ~Kim, DN~Basov, and
  RD~Averitt.
\newblock {Conductivity dynamics in the correlated metallic state of V2O3}.
\newblock In \emph{Lasers and Electro-Optics, 2009 and 2009 Conference on
  Quantum electronics and Laser Science Conference. CLEO/QELS 2009. Conference
  on}, volume~4, pages 1--2. IEEE, 2009.
\newblock URL
  \url{http://ieeexplore.ieee.org/xpls/abs\_all.jsp?arnumber=5225800}.

\bibitem[Zylbersztejn and Mott(1975)]{Zylbersztejn1975}
A.~Zylbersztejn and N.F. Mott.
\newblock {Metal-insulator transition in vanadium dioxide}.
\newblock \emph{Physical Review B}, 11\penalty0 (11):\penalty0 4383, 1975.

\bibitem[Kim et~al.(2004)Kim, Chae, Youn, Maeng, Kim, Kang, and Lim]{Kim2004}
Hyun-Tak Kim, B~G Chae, D~H Youn, S~L Maeng, G~Kim, K~Y Kang, and Y~S Lim.
\newblock {Mechanism and observation of Mott transition in VO 2 based two- and
  three-terminal devices}.
\newblock \emph{New Journal of Physics}, 6:\penalty0 52--70, 2004.

\bibitem[Rice et~al.(1994)Rice, Launois, and Pouget]{Rice1994b}
T.M. Rice, H.~Launois, and J.P. Pouget.
\newblock {Comment on "VO2: Peierls or Mott-Hubbard? A View from Band Theory"}.
\newblock \emph{Physical Review Letters}, 73\penalty0 (22):\penalty0
  9007--9007, 1994.

\bibitem[Wentzcovitch et~al.(1994)Wentzcovitch, Schulz, and
  Allen]{Wentzcovitch1994}
Renata~M Wentzcovitch, Werner~W Schulz, and Philip~B Allen.
\newblock {VO2: Peierls or Mott-Hubbard? A View from Band Theory}.
\newblock \emph{Physical Review Letters}, 72\penalty0 (21):\penalty0
  3389--3392, 1994.

\bibitem[Cavalleri et~al.(2004)Cavalleri, Dekorsy, Chong, Kieffer, and
  Schoenlein]{Cavalleri2004}
Andrea Cavalleri, Th. Dekorsy, H.~Chong, J.~Kieffer, and R.~Schoenlein.
\newblock {Evidence for a structurally-driven insulator-to-metal transition in
  VO2: A view from the ultrafast timescale}.
\newblock \emph{Physical Review B}, 70\penalty0 (16):\penalty0 3--6, October
  2004.
\newblock ISSN 1098-0121.
\newblock \doi{10.1103/PhysRevB.70.161102}.
\newblock URL \url{http://link.aps.org/doi/10.1103/PhysRevB.70.161102}.

\bibitem[Cavalleri et~al.(2001)Cavalleri, T\'{o}th, Siders, Squier, R\'{a}ksi,
  Forget, and Kieffer]{Cavalleri2001}
Andrea Cavalleri, Cs. T\'{o}th, C~W Siders, J~A Squier, F~R\'{a}ksi, P~Forget,
  and J~C Kieffer.
\newblock {Femtosecond structural dynamics in VO 2 during an ultrafast
  solid-solid phase transition}.
\newblock \emph{PRL}, 87237401:\penalty0 1--4, 2001.

\bibitem[Rini et~al.(2005)Rini, Cavalleri, Schoenlein, L\'{o}pez, Feldman,
  Haglund, Boatner, and Haynes]{Rini2005}
Matteo Rini, Andrea Cavalleri, Robert~W Schoenlein, Ren\'{e} L\'{o}pez,
  Leonard~C Feldman, Richard~F Haglund, Lynn~A Boatner, and Tony~E Haynes.
\newblock {Photoinduced phase transition in VO 2 nanocrystals : control of
  surface-plasmon resonance ultrafast}.
\newblock \emph{Optics Letters}, 30\penalty0 (5):\penalty0 1--3, 2005.

\bibitem[Lysenko et~al.(2006)Lysenko, Rua, Vikhnin, Jimenez, Fernandez, and
  Liu]{Lysenko2006}
S~Lysenko, AJ~Rua, V~Vikhnin, J~Jimenez, F~Fernandez, and H~Liu.
\newblock {Light-induced ultrafast phase transitions in VO2 thin film}.
\newblock \emph{Applied surface science}, 252\penalty0 (15):\penalty0
  5512--5515, 2006.
\newblock \doi{10.1016/j.apsusc.2005.12.137}.
\newblock URL
  \url{http://linkinghub.elsevier.com/retrieve/pii/S0169433205018027}.

\bibitem[Lopez et~al.(2004)Lopez, Boatner, Haynes, Haglund, and
  Feldman]{Lopez2004}
René Lopez, L.~a. Boatner, T.~E. Haynes, R.~F. Haglund, and L.~C. Feldman.
\newblock {Switchable reflectivity on silicon from a composite VO[sub
  2]-SiO[sub 2] protecting layer}.
\newblock \emph{Applied Physics Letters}, 85\penalty0 (8):\penalty0 1410, 2004.
\newblock ISSN 00036951.
\newblock \doi{10.1063/1.1784546}.
\newblock URL \url{http://link.aip.org/link/APPLAB/v85/i8/p1410/s1\&Agg=doi}.

\bibitem[Driscoll et~al.(2008)Driscoll, Palit, Qazilbash, Brehm, Keilmann,
  Chae, Yun, Kim, Jokerst, Smith, and Basov]{Driscoll2008}
Tom Driscoll, S~Palit, Mumtaz~M. Qazilbash, Markus Brehm, Fritz Keilmann,
  Byung-Gyu Chae, Sun-Jin Yun, Hyun-Tak Kim, Nan~Marie Jokerst, David~R. Smith,
  and Dimitri~N. Basov.
\newblock {Dynamic tuning of an infrared hybrid-metamaterial resonance using
  vanadium dioxide}.
\newblock \emph{Applied Physics Letters}, 93\penalty0 (024101), 2008.
\newblock \doi{10.1063/1.2956675}.
\newblock URL
  \url{http://ieeexplore.ieee.org/xpls/abs\_all.jsp?arnumber=4838777}.

\bibitem[Driscoll et~al.(2009{\natexlab{a}})Driscoll, Kim, Chae, Kim, Lee,
  Jokerst, Palit, Smith, {Di Ventra}, and Basov]{Driscoll2009b}
Tom Driscoll, Hyun-Tak Kim, Byung-Gyu Chae, Bong-Jun Kim, Yong-Wook Lee,
  Nan~Marie Jokerst, S~Palit, David~R. Smith, Massimiliano {Di Ventra}, and
  Dimitri~N. Basov.
\newblock {Memory metamaterials.}
\newblock \emph{Science (New York, N.Y.)}, 325\penalty0 (5947):\penalty0
  1518--21, September 2009{\natexlab{a}}.
\newblock ISSN 1095-9203.
\newblock \doi{10.1126/science.1176580}.
\newblock URL \url{http://www.ncbi.nlm.nih.gov/pubmed/19696311}.

\bibitem[Dicken et~al.(2009)Dicken, Aydin, Pryce, Sweatlock, Boyd, Walavalkar,
  Ma, and Atwater]{Dicken2009}
Matthew~J Dicken, Koray Aydin, Imogen~M Pryce, Luke~A Sweatlock, M~Boyd, Sameer
  Walavalkar, James Ma, and Harry~A Atwater.
\newblock {Frequency tunable near-infrared metamaterials based on VO 2 phase
  transition}.
\newblock \emph{Optics Express}, 17\penalty0 (20):\penalty0 295--298, 2009.

\bibitem[Kim et~al.(2006)Kim, Lee, Chae, Yun, Oh, Lim, and Kim]{Kim2006}
B.J. Kim, Y.W. Lee, B.G. Chae, S.J. Yun, S.Y. Oh, Y.S. Lim, and Hyun-Tak Kim.
\newblock {Temperature dependence of Mott transition in VO$\backslash$\_2 and
  programmable critical temperature sensor}.
\newblock \emph{Arxiv preprint cond-mat/0609033}, 009033v1, 2006.
\newblock URL \url{http://arxiv.org/abs/cond-mat/0609033}.

\bibitem[Driscoll et~al.(2010)Driscoll, Quinn, Klein, Kim, Kim, Pershin, {Di
  Ventra}, and Basov]{Driscoll2010a}
Tom Driscoll, J.~Quinn, S.~Klein, Hyun-Tak Kim, B.~J. Kim, Yuriy~V. Pershin,
  Massimiliano {Di Ventra}, and D.~N. Basov.
\newblock {Memristive adaptive filters}.
\newblock \emph{Applied Physics Letters}, 97\penalty0 (9):\penalty0 093502,
  2010.
\newblock ISSN 00036951.
\newblock \doi{10.1063/1.3485060}.
\newblock URL \url{http://link.aip.org/link/APPLAB/v97/i9/p093502/s1\&Agg=doi}.

\bibitem[Driscoll et~al.(2009{\natexlab{b}})Driscoll, Kim, Chae, {Di Ventra},
  and Basov]{Driscoll2009d}
Tom Driscoll, Hyun-Tak Kim, B.G. Chae, Massimiliano {Di Ventra}, and DN~Basov.
\newblock {Phase-transition driven memristive system}.
\newblock \emph{Applied Physics Letters}, 95\penalty0 (4):\penalty0 043503,
  2009{\natexlab{b}}.
\newblock URL \url{http://link.aip.org/link/?APPLAB/95/043503/1}.

\bibitem[Pershin(2011)]{Pershin2011}
YV~Pershin.
\newblock {Memory effects in complex materials and nanoscale systems}.
\newblock \emph{Advances in Physics}, 00\penalty0 (00):\penalty0 1--71, 2011.
\newblock \doi{10.1080/0001873YYxxxxxxxx}.
\newblock URL
  \url{http://www.tandfonline.com/doi/abs/10.1080/00018732.2010.544961}.

\bibitem[Crunteanu et~al.(2010)Crunteanu, Givernaud, Leroy, Mardivirin,
  Champeaux, Orlianges, Catherinot, and Blondy]{Crunteanu2010}
Aurelian Crunteanu, Julien Givernaud, Jonathan Leroy, David Mardivirin, Corinne
  Champeaux, Jean-Christophe Orlianges, Alain Catherinot, and Pierre Blondy.
\newblock {Voltage- and current-activated metal–insulator transition in VO 2
  -based electrical switches: a lifetime operation analysis}.
\newblock \emph{Science and Technology of Advanced Materials}, 11\penalty0
  (6):\penalty0 065002, December 2010.
\newblock ISSN 1468-6996.
\newblock \doi{10.1088/1468-6996/11/6/065002}.
\newblock URL
  \url{http://stacks.iop.org/1468-6996/11/i=6/a=065002?key=crossref.3559a62a92655ee932bf6fc313bbb47a}.

\bibitem[Lee et~al.(2008)Lee, Kim, Lim, Yun, Choi, Chae, Kim, and
  Kim]{Lee2008a}
Y.W. Lee, B.J. Kim, J.W. Lim, S.J. Yun, Sungyoul Choi, B.G. Chae, G.~Kim, and
  Hyun-Tak Kim.
\newblock {Metal-insulator transition-induced electrical oscillation in
  vanadium dioxide thin film}.
\newblock \emph{Applied Physics Letters}, 92\penalty0 (16):\penalty0 162903,
  2008.
\newblock \doi{10.1063/1.2911745}.
\newblock URL \url{http://link.aip.org/link/?APPLAB/92/162903/1}.

\bibitem[Sakai(2008)]{Sakai2008}
Joe Sakai.
\newblock {High-efficiency voltage oscillation in VO[sub 2] planer-type
  junctions with infinite negative differential resistance}.
\newblock \emph{Journal of Applied Physics}, 103\penalty0 (10):\penalty0
  103708, 2008.
\newblock ISSN 00218979.
\newblock \doi{10.1063/1.2930959}.
\newblock URL
  \url{http://link.aip.org/link/JAPIAU/v103/i10/p103708/s1\&Agg=doi}.

\bibitem[Kim et~al.(2010{\natexlab{a}})Kim, Kim, Choi, Chae, Lee, Driscoll,
  Qazilbash, and Basov]{Kim2010b}
Hyun-Tak Kim, Bong-Jun Kim, Sungyoul Choi, Byung-Gyu Chae, Yong~Wook Lee, Tom
  Driscoll, Mumtaz~M. Qazilbash, and D.~N. Basov.
\newblock {Electrical oscillations induced by the metal-insulator transition in
  VO[sub 2]}.
\newblock \emph{Journal of Applied Physics}, 107\penalty0 (2):\penalty0 023702,
  2010{\natexlab{a}}.
\newblock ISSN 00218979.
\newblock \doi{10.1063/1.3275575}.
\newblock URL
  \url{http://link.aip.org/link/JAPIAU/v107/i2/p023702/s1\&Agg=doi}.

\bibitem[Kim et~al.(2010{\natexlab{b}})Kim, Seo, Lee, Choi, and Kim]{Kim2010c}
B.J. Kim, Giwan Seo, Y.W. Lee, Sungyoul Choi, and Hyun-Tak Kim.
\newblock {Linear Characteristics of a Metal–Insulator Transition Voltage and
  Oscillation Frequency in VO2 Devices}.
\newblock \emph{Electron Device Letters, IEEE}, 31\penalty0 (11):\penalty0
  1314--1316, 2010{\natexlab{b}}.
\newblock URL
  \url{http://ieeexplore.ieee.org/xpls/abs\_all.jsp?arnumber=5595090}.

\bibitem[Sharoni et~al.(2008)Sharoni, Ram$\backslash$'$\backslash$irez, and
  Schuller]{Sharoni2008}
Amos Sharoni, J.G. Ram$\backslash$'$\backslash$irez, and I.K. Schuller.
\newblock {Multiple avalanches across the metal-insulator transition of
  vanadium oxide nanoscaled junctions}.
\newblock \emph{Physical review letters}, 101\penalty0 (2):\penalty0 26404,
  2008.
\newblock \doi{10.1103/PhysRevLett.101.026404}.
\newblock URL \url{http://link.aps.org/doi/10.1103/PhysRevLett.101.026404}.

\bibitem[Ram$\backslash$'$\backslash$irez
  et~al.(2009)Ram$\backslash$'$\backslash$irez, Sharoni, Dubi, Gomez, and
  Schuller]{Ramirez2009}
J.G. Ram$\backslash$'$\backslash$irez, Amos Sharoni, Y~Dubi, ME~Gomez, and I.K.
  Schuller.
\newblock {First-order reversal curve measurements of the metal-insulator
  transition in VO$_2$: Signatures of persistent metallic domains}.
\newblock \emph{Physical Review B}, 79\penalty0 (23):\penalty0 235110, 2009.
\newblock \doi{10.1103/PhysRevB.79.235110}.
\newblock URL \url{http://prb.aps.org/abstract/PRB/v79/i23/e235110}.

\bibitem[{Di Ventra} et~al.(2009){Di Ventra}, Pershin, and Chua]{DiVentra2009}
Massimiliano {Di Ventra}, Yuriy~V. Pershin, and Leon~O. Chua.
\newblock {Circuit elements with memory: memristors, memcapacitors and
  meminductors}.
\newblock \emph{Proceedings of the IEEE}, 97\penalty0 (8):\penalty0 1371--1372,
  January 2009.
\newblock \doi{10.1109/JPROC.2009.2021077}.
\newblock URL \url{http://arxiv.org/abs/0901.3682}.

\bibitem[Rozen et~al.(2006)Rozen, Lopez, Haglund, and Feldman]{Rozen2006}
John Rozen, René Lopez, Richard~F. Haglund, and Leonard~C. Feldman.
\newblock {Two-dimensional current percolation in nanocrystalline vanadium
  dioxide films}.
\newblock \emph{Applied Physics Letters}, 88\penalty0 (8):\penalty0 081902,
  2006.
\newblock ISSN 00036951.
\newblock \doi{10.1063/1.2175490}.
\newblock URL \url{http://link.aip.org/link/APPLAB/v88/i8/p081902/s1\&Agg=doi}.

\bibitem[Dai et~al.(2008)Dai, Wang, Huang, and Yi]{Dai2008}
Jun Dai, Xingzhi Wang, Ying Huang, and Xinjian Yi.
\newblock {Modeling of temperature-dependent resistance in micro- and
  nanopolycrystalline VO[sub 2] thin films with random resistor networks}.
\newblock \emph{Optical Engineering}, 47\penalty0 (3):\penalty0 033801, 2008.
\newblock ISSN 00913286.
\newblock \doi{10.1117/1.2894146}.
\newblock URL \url{http://link.aip.org/link/OPEGAR/v47/i3/p033801/s1\&Agg=doi}.

\bibitem[Pan et~al.(2004)Pan, Zhong, Wang, Liu, Li, Chen, and Lu]{Pan2004}
Mei Pan, Hongmei Zhong, Shaowei Wang, Jie Liu, Zhifeng Li, Xiaoshuang Chen, and
  Wei Lu.
\newblock {Properties of VO 2 thin film prepared with precursor VO ( acac ) 2}.
\newblock \emph{Journal of Crystal Growth}, 265:\penalty0 121--126, 2004.
\newblock \doi{10.1016/j.jcrysgro.2003.12.065}.

\bibitem[Mlyuka and Kivaisi(2006)]{Mlyuka2006}
N.~R. Mlyuka and R.~T. Kivaisi.
\newblock {Correlation between optical, electrical and structural properties of
  vanadium dioxide thin films}.
\newblock \emph{Journal of Materials Science}, 41\penalty0 (17):\penalty0
  5619--5624, June 2006.
\newblock ISSN 0022-2461.
\newblock \doi{10.1007/s10853-006-0261-y}.
\newblock URL
  \url{http://www.springerlink.com/index/10.1007/s10853-006-0261-y}.

\bibitem[Kim et~al.(2010{\natexlab{c}})Kim, Ko, Frenzel, Ramanathan, and
  Hoffman]{Kim2010d}
Jeehoon Kim, Changhyun Ko, Alex Frenzel, Shriram Ramanathan, and Jennifer~E.
  Hoffman.
\newblock {Nanoscale imaging and control of resistance switching in VO[sub 2]
  at room temperature}.
\newblock \emph{Applied Physics Letters}, 96\penalty0 (21):\penalty0 213106,
  2010{\natexlab{c}}.
\newblock ISSN 00036951.
\newblock \doi{10.1063/1.3435466}.
\newblock URL
  \url{http://link.aip.org/link/APPLAB/v96/i21/p213106/s1\&Agg=doi}.

\bibitem[Qazilbash et~al.(2011)Qazilbash, Tripathi, Schafgans, Kim, Kim, Cai,
  Holt, Maser, Keilmann, Shpyrko, and Basov]{Qazilbash2011}
Mumtaz~M. Qazilbash, a.~Tripathi, a.~Schafgans, Bong-Jun Kim, Hyun-Tak Kim,
  Zhonghou Cai, M.~Holt, J.~Maser, F.~Keilmann, O.~Shpyrko, and D.~Basov.
\newblock {Nanoscale imaging of the electronic and structural transitions in
  vanadium dioxide}.
\newblock \emph{Physical Review B}, 83\penalty0 (16):\penalty0 1--7, April
  2011.
\newblock ISSN 1098-0121.
\newblock \doi{10.1103/PhysRevB.83.165108}.
\newblock URL \url{http://link.aps.org/doi/10.1103/PhysRevB.83.165108}.

\bibitem[Frenzel et~al.(2009)Frenzel, Qazilbash, Brehm, Chae, Kim, Kim,
  Balatsky, Keilmann, and Basov]{Frenzel2009}
Alex. Frenzel, Mumtaz~M. Qazilbash, M.~Brehm, Byung-Gyu Chae, Bong-Jun Kim,
  Hyun-Tak Kim, A.~Balatsky, F.~Keilmann, and D.~Basov.
\newblock {Inhomogeneous electronic state near the insulator-to-metal
  transition in the correlated oxide VO2}.
\newblock \emph{Physical Review B}, 80\penalty0 (11):\penalty0 1--7, September
  2009.
\newblock ISSN 1098-0121.
\newblock \doi{10.1103/PhysRevB.80.115115}.
\newblock URL \url{http://link.aps.org/doi/10.1103/PhysRevB.80.115115}.

\bibitem[Poklonski et~al.(2006)Poklonski, Kocherzhenko, Benediktovitch,
  Mitsianok, and Zaitsev]{Poklonski2006}
N.~a. Poklonski, a.~a. Kocherzhenko, a.~I. Benediktovitch, V.~V. Mitsianok, and
  a.~M. Zaitsev.
\newblock {Simulation of dc conductance of two-dimensional heterogeneous
  system: application to carbon wires made by ion irradiation on
  polycrystalline diamond}.
\newblock \emph{Physica Status Solidi (B)}, 243\penalty0 (6):\penalty0
  1212--1218, May 2006.
\newblock ISSN 0370-1972.
\newblock \doi{10.1002/pssb.200541079}.
\newblock URL \url{http://doi.wiley.com/10.1002/pssb.200541079}.

\bibitem[Ruzmetov et~al.(2009)Ruzmetov, Gopalakrishnan, Deng, Narayanamurti,
  and Ramanathan]{Ruzmetov2009}
Dmitry Ruzmetov, Gokul Gopalakrishnan, Jiangdong Deng, V.~Narayanamurti, and
  Shriram Ramanathan.
\newblock {Electrical triggering of metal-insulator transition in nanoscale
  vanadium oxide junctions}.
\newblock \emph{Journal of Applied Physics}, 106\penalty0 (8):\penalty0
  083702--083702, 2009.
\newblock \doi{10.1063/1.3245338}.
\newblock URL
  \url{http://ieeexplore.ieee.org/xpls/abs\_all.jsp?arnumber=5292009}.

\bibitem[Stefanovich et~al.(2000)Stefanovich, Pergament, and
  Stefanovich]{Stefanovich2000a}
G.~Stefanovich, A~Pergament, and D~Stefanovich.
\newblock {Electrical switching and Mott transition in VO2}.
\newblock \emph{Journal of Physics: Condensed Matter}, 12:\penalty0 8837, 2000.
\newblock URL \url{http://iopscience.iop.org/0953-8984/12/41/310}.

\bibitem[Gopalakrishnan et~al.(2009)Gopalakrishnan, Ruzmetov, and
  Ramanathan]{Gopalakrishnan2009a}
Gokul Gopalakrishnan, D.~Ruzmetov, and Shriram Ramanathan.
\newblock {On the triggering mechanism for the metal–insulator transition in
  thin film VO 2 devices: electric field versus thermal effects}.
\newblock \emph{Journal of materials science}, 44\penalty0 (19):\penalty0
  5345--5353, 2009.
\newblock \doi{10.1007/s10853-009-3442-7}.
\newblock URL \url{http://www.springerlink.com/index/7U24554G53291604.pdf}.

\bibitem[{Di Ventra}(2008)]{DiVentra2008}
Massimiliano {Di Ventra}.
\newblock \emph{{Electrical transport in nanoscale systems}}.
\newblock Cambridge University Press, Cambridge, 2008.

\bibitem[Chudnovskiy et~al.(2002)Chudnovskiy, Luryi, and
  Spivak]{Chudnovskiy2002a}
Feliks Chudnovskiy, Serge Luryi, and Boris Spivak.
\newblock {Switching device based on first-order metal- insulator transition
  induced by external electric field}.
\newblock \emph{Future Trends in Microelectronics: the Nano Millennium}, pages
  148--155, 2002.

\bibitem[Qazilbash et~al.(2008{\natexlab{b}})Qazilbash, Li, Podzorov, Brehm,
  Keilmann, Chae, Kim, and Basov]{Qazilbash2008}
Mumtaz~M. Qazilbash, Z.~Q. Li, V.~Podzorov, M.~Brehm, F.~Keilmann, B.~G. Chae,
  Hyun-Tak Kim, and D.~N. Basov.
\newblock {Electrostatic modification of infrared response in gated structures
  based on VO[sub 2]}.
\newblock \emph{Applied Physics Letters}, 92\penalty0 (24):\penalty0 241906,
  2008{\natexlab{b}}.
\newblock ISSN 00036951.
\newblock \doi{10.1063/1.2939434}.
\newblock URL
  \url{http://link.aip.org/link/APPLAB/v92/i24/p241906/s1\&Agg=doi}.

\bibitem[Wu et~al.(2011)Wu, Whittaker, Banerjee, and Sambandamurthy]{Wu2011}
Tai-Lung Wu, Luisa Whittaker, Sarbajit Banerjee, and G.~Sambandamurthy.
\newblock {Temperature and voltage driven tunable metal-insulator transition in
  individual W$_x$V$_{1-x}$O$_2$ nanowires}.
\newblock \emph{Physical Review B}, 83\penalty0 (7):\penalty0 2--5, February
  2011.
\newblock ISSN 1098-0121.
\newblock \doi{10.1103/PhysRevB.83.073101}.
\newblock URL \url{http://link.aps.org/doi/10.1103/PhysRevB.83.073101}.

\bibitem[Arcangeletti et~al.(2007)Arcangeletti, Baldassarre, {Di Castro}, Lupi,
  Malavasi, Marini, Perucchi, and Postorino]{Arcangeletti2007}
E.~Arcangeletti, L.~Baldassarre, D.~{Di Castro}, S.~Lupi, L.~Malavasi,
  C.~Marini, a.~Perucchi, and P.~Postorino.
\newblock {Evidence of a Pressure-Induced Metallization Process in Monoclinic
  VO2}.
\newblock \emph{Physical Review Letters}, 98\penalty0 (19):\penalty0 1--4, May
  2007.
\newblock ISSN 0031-9007.
\newblock \doi{10.1103/PhysRevLett.98.196406}.
\newblock URL \url{http://link.aps.org/doi/10.1103/PhysRevLett.98.196406}.

\bibitem[Wei et~al.(2009)Wei, Wang, Chen, and Cobden]{Wei2009}
Jiang Wei, Zenghui Wang, Wei Chen, and D.H. Cobden.
\newblock {New aspects of the metal–insulator transition in single-domain
  vanadium dioxide nanobeams}.
\newblock \emph{Nature Nanotechnology}, 4\penalty0 (7):\penalty0 420--424,
  2009.
\newblock \doi{10.1038/NNANO.2009.141}.
\newblock URL
  \url{http://www.nature.com/nnano/journal/v4/n7/abs/nnano.2009.141.html}.

\bibitem[Yao et~al.(2010)Yao, Zhang, Sun, Liu, Huang, Xie, Wu, Yuan, Zhang, Wu,
  Pan, Hu, Wu, Liu, and Wei]{Yao2010a}
Tao Yao, Xiaodong Zhang, Zhihu Sun, Shoujie Liu, Yuanyuan Huang, Yi~Xie,
  Changzheng Wu, Xun Yuan, Wenqing Zhang, Ziyu Wu, Guoqiang Pan, Fengchun Hu,
  Lihui Wu, Qinghua Liu, and Shiqiang Wei.
\newblock {Understanding the Nature of the Kinetic Process in a VO2
  Metal-Insulator Transition}.
\newblock \emph{Physical Review Letters}, 105\penalty0 (22):\penalty0 2--5,
  November 2010.
\newblock ISSN 0031-9007.
\newblock \doi{10.1103/PhysRevLett.105.226405}.
\newblock URL \url{http://link.aps.org/doi/10.1103/PhysRevLett.105.226405}.

\bibitem[Kim et~al.(2007)Kim, Lee, Chae, Yun, Oh, Kim, and Lim]{Kim2007}
Bong-Jun Kim, Yong~Wook Lee, Byung-Gyu Chae, Sun~Jin Yun, Soo-Young Oh,
  Hyun-Tak Kim, and Yong-Sik Lim.
\newblock {Temperature dependence of the first-order metal-insulator transition
  in VO[sub 2] and programmable critical temperature sensor}.
\newblock \emph{Applied Physics Letters}, 90\penalty0 (2):\penalty0 023515,
  2007.
\newblock ISSN 00036951.
\newblock \doi{10.1063/1.2431456}.
\newblock URL \url{http://link.aip.org/link/APPLAB/v90/i2/p023515/s1\&Agg=doi}.

\bibitem[Kim et~al.(2008)Kim, Lee, Choi, Lim, Yun, Kim, Shin, and
  Yun]{Kim2008b}
Bong-Jun Kim, Yong Lee, Sungyeoul Choi, Jung-Wook Lim, Sun Yun, Hyun-Tak Kim,
  Tae-Ju Shin, and Hwa-Sick Yun.
\newblock {Micrometer x-ray diffraction study of VO2 films: Separation between
  metal-insulator transition and structural phase transition}.
\newblock \emph{Physical Review B}, 77\penalty0 (23):\penalty0 1--5, June 2008.
\newblock ISSN 1098-0121.
\newblock \doi{10.1103/PhysRevB.77.235401}.
\newblock URL \url{http://link.aps.org/doi/10.1103/PhysRevB.77.235401}.

\bibitem[Kim et~al.(2005)Kim, Chae, Youn, Kim, Kang, Lee, Kim, and
  Lim]{Kim2005}
Hyun-Tak Kim, Byung-Gyu Chae, Doo-Hyeb Youn, Gyungock Kim, Kwang-Yong Kang,
  Seung-Joon Lee, Kwan Kim, and Yong-Sik Lim.
\newblock {Raman study of electric-field-induced first-order metal-insulator
  transition in VO[sub 2]-based devices}.
\newblock \emph{Applied Physics Letters}, 86\penalty0 (24):\penalty0 242101,
  2005.
\newblock ISSN 00036951.
\newblock \doi{10.1063/1.1941478}.
\newblock URL
  \url{http://link.aip.org/link/APPLAB/v86/i24/p242101/s1\&Agg=doi}.

\end{thebibliography}

\end{document}